\documentclass{jetpl}

\twocolumn
\lat

\title{The quantum Talbot effect for a chain of partially correlated Bose--Einstein condensates}

\rtitle{The quantum Talbot effect for a chain of partially correlated Bose--Einstein condensates}
\sodtitle{The quantum Talbot effect for a chain of partially correlated Bose--Einstein condensates}
\author{V.\,B. Makhalov, A.\,V. Turlapov\/\thanks{turlapov@appl.sci-nnov.ru}}
\rauthor{V.\,B. Makhalov, A.\,V. Turlapov}
\sodauthor{V.\,B. Makhalov, A.\,V. Turlapov}
\address{Institute of applied physics, RAS, 603950, Nizhniy Novgorod, Russia}
\dates{\today}{*}

\abstract{The matter-wave interference picture, which appears within the quantum Talbot effect, changes qualitatively in response to even a small randomness in the phases of the sources. The spatial spectrum acquires peaks which are absent in the case of equal phases. The mathematic model of the effect is presented. The manifestations of the phase randomness in experiments with atomic and molecular Bose condensates, is discussed. Thermometry based on observable consequences of phase randomness is proposed.}

\begin{document}\maketitle

\section{Introduction}

In the classical optics, self-imaging of a field periodically distributed in the initial plane is referred to as the Talbot effect~\cite{Talbot}. Similar phenomena have been later seen for acoustic waves~\cite{AcousticTalbot1985,USoundTalbot2017eng}, wave functions of atoms~\cite{PritchardTalbot1995} and electrons~\cite{DenisovTalbot1996,DenisovTalbot1999}, plasmons~\cite{PlasmonTalbot2009}, and spin waves~\cite{SpinwaveTalbot2012}. In optics, the initial distribution of the field is reproduced in the paraxial approximation. In quantum mechanics, the effect is exact for a periodic wave function $\psi(z+d)=\psi(z)$ and free-particle Hamiltonian $\hat H=\hat p^2/2M$. In the quantum case, the spatial propagation of the wave front is not necessary because the wave function may recover at the same location.

In a quantum many-body system limited in the $x$ and $y$ directions, the phase fluctuations complicate or prohibit fulfillment of the Talbot-effect initial condition, namely the formation of an exactly periodic along $z$ chain of sources. In a Bose--Einstein condensate (BEC) extended along $z$, the phase fluctuates down to temperatures well below the condensation temperature~\cite{ShlyapnikovElongatedBECFluct2001}. In an infinitely long chain of superconductors or BECs,  long-wavelength phase fluctuations are present even at zero temperature~\cite{BradleyDoniach1984} making phases of adjacent sources at least partially uncorrelated. Any small phase fluctuation qualitatively alters the interference picture in comparison to the Talbot effect~\cite{RandomPhaseInteference2019}. In the spatial-distribution spectrum, the peak with wave vector $k=2\pi/d$ corresponding to the initial-modulation period is partially preserved on one hand, while on the other hand, peaks with wave vectors $k<2\pi/d$ appear, as observed in experiment with a BEC chain~\cite{RandomPhaseInteference2019}.

This paper is devoted to interference of a chain of BECs whose phases are only partially correlated. In section~\ref{sec:InterferenceModel}, an interference model is presented for the limiting cases of fully correlated and fully uncorrelated phases and a numeric simulation is performed for the intermediated case of partially correlated phases. Section~\ref{sec:InterferenceExperiments} is devoted to experiments where BECs with fluctuating phases interfere. In section~\ref{sec:Thermometry}, thermometry for a BEC chain is proposed relying on the contribution of the phase fluctuation into the interference-fringe spatial spectrum. The conclusion is given in section~\ref{sec:Concl}.

\section{Interference model for a chain of sources with unrestricted phases}\label{sec:InterferenceModel}

\begin{figure}
\begin{center}\includegraphics[width=\linewidth]{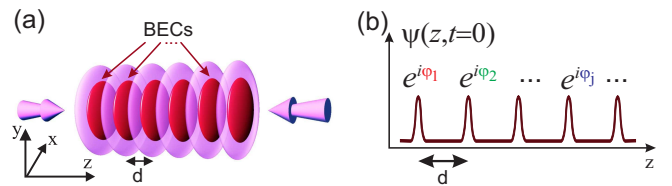}\end{center}
\caption{FIGURE~1. (a) BECs in an optical lattice right before the lattice extinction and the onset of the BEC expansion and interference. The BECs are shown in dark red, while the lattice is light purple. (b) The initial wave function of the interfering condensates: The absolute value is periodic, while the adjacent-condensate phase difference is unrestricted.}
\label{fig:InitialCondition}
\end{figure}

The chain of sources is modeled by sum of localized wave functions
\begin{equation}
\psi(z,t=0)=\sqrt{\frac N{\sigma\sqrt{2\pi}}} \sum_{j=1}^Ke^{-(z-jd)^2/4\sigma^2}\,e^{i\varphi_j},
\label{eq:PsiInitial}
\end{equation}
where $\sigma\ll d$ and the chain is long, i.~e., $K\rightarrow\infty$. The corresponding density is periodic,  $|\psi(z+d)|^2=|\psi(z)|^2$. Phases $\varphi_j$ are generally unrestricted. Wave function~(\ref{eq:PsiInitial}) corresponds to the chain of BECs in optical lattice shown in figure~\ref{fig:InitialCondition}a, while wave function~(\ref{eq:PsiInitial}) by itself is depicted in figure~\ref{fig:InitialCondition}b.

In the case of equal phases $\varphi_j=\text{inv}(j)$ wave function~(\ref{eq:PsiInitial}) corresponds to the initial condition of the Talbot effect.
By applying the free-space evolution operator $\exp(-i\hat p^2t/(2M\hbar))$ for mass $M$ bosons,
the wave function may be propagated to any time $t$. One may see that $\psi(z,t)$ is spatially periodic with period $d$ at any $t$, while at the times that are integer multiples of Talbot time $T_d\equiv Md^2/\pi\hbar$
the initial wave function reestablishes, i.~e., $\psi(z,t=nT_d)=\psi(z,t=0)$ for $n\in\mathbb{N}$.

\begin{figure}[htb!]
\begin{center}
\includegraphics[width=0.8\linewidth]{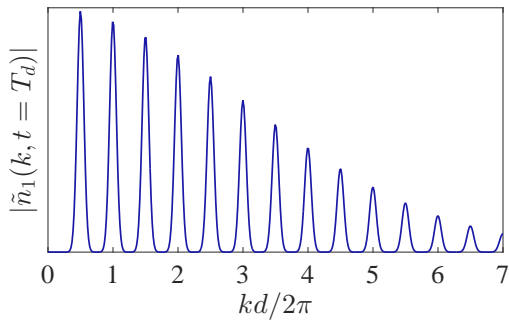}
\end{center}
\caption{FIGURE~2. Absolute value of spectrum~(\ref{eq:TalbotDensitySpectrum}) at $t=T_d$. The principal harmonic at $k=\pi/d$ corresponds to spatial period $2d$.}
\label{fig:FresnelSpectrum}
\end{figure}

The opposite limit corresponds to condensate phases $\varphi_j$ which are fully random with respect to each other. In this case, the free space evolution of wave function~(\ref{eq:PsiInitial}) may again be calculated. This in turn allows calculation of the density-distribution spectrum
\begin{equation}
\tilde{n}_1(k,t)=\int\limits_{-\infty}^\infty|\psi(z,t)|^2e^{-ikz}dz.
\end{equation}
In the long-chain limit $K\rightarrow\infty$, spectrum $\tilde{n}_1(k,t)$ takes form~\cite{RandomPhaseInteference2019}
\begin{align}
\tilde{n}_1(k,t)&\propto\frac{2\pi}d\delta(k) + \frac{\sqrt{\pi K}}2 e^{-k^2\sigma^2/2}\times\nonumber\\ &\sum_{j=-\infty,\,j\neq0}^{\infty} e^{-(j-kdt/T_d\pi)^2 d^2/8\sigma^2} e^{i\varphi_j'},
\label{eq:TalbotDensitySpectrum}
\end{align}
where $\varphi_j'$ are random phases satisfying $\varphi_j'=-\varphi_{-j}'$. Owing to condition $\sigma\ll d$, spectrum~(\ref{eq:TalbotDensitySpectrum}) is a sequence of peaks as seen, for example, in figure~\ref{fig:FresnelSpectrum}.
At general instant $t$ the spectrum is a sum of harmonics with wave vectors $k=j\pi T_d/(td)$ with $j\in\mathbb{N}$, while one-dimensional density distribution $n_1(z,t)$ has spatial period
\begin{equation}
\frac{2\pi}k=d\frac{2t}{T_d}.\label{eq:RandomPhasePeriod}
\end{equation}
The period is linearly growing with time. Interestingly, $n_1(z,t)$ is periodic despite the phase randomness. The value of the spatial period makes qualitative difference with the Talbot effect, where longer than $d$ principal periods are absent.

\begin{figure}[htb!]
\begin{center}
\includegraphics[width=0.8\linewidth]{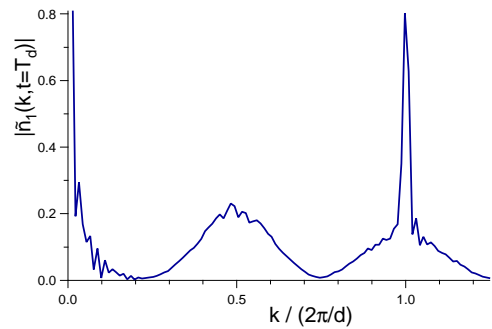}
\end{center}
\caption{FIGURE~3. Numerically calculated absolute value of spectrum $|\tilde{n}_1(k,t=T_d)|$ for partial correlation. The correlation degree $\langle\cos(\varphi_{j+1}-\varphi_j)\rangle=0.57$ is chosen in such a way that for the peaks at $k=2\pi/d$ the ratio of the narrow-peak height to the wide-peak height is the same as in the experimental data of figure~\ref{fig:TurlapovTalbot}c.
}
\label{fig:PartialCorrSpectrum}
\end{figure}

For the intermediate case of partially correlated phases, an analytic solution is unknown so far. The evolution of wave function~(\ref{eq:PsiInitial}) may be found numerically. The spatial-spectrum absolute value $|\tilde n_1(k)|$ computed for $t=T_d$ is shown in figure~\ref{fig:PartialCorrSpectrum}.
This spectrum combines features of both the Talbot effect and uncorrelated-source interference. The narrow peak at $k=2\pi/d$ corresponds to partially preserved Talbot effect, while the wide peaks centered near $k=\pi/d$ and $k=2\pi/d$ are due to the partial phase disorder. The calculation is done for $K=51$ BECs, ratio $\sigma/d=0.1$, correlation degree $\langle\cos(\varphi_{j+1}-\varphi_j)\rangle=0.57$. The spectrum absolute value $|\tilde{n}_1(k,t=T_d)|$ is averaged over 100 repetitions.

\section{Experimental manifestation of the phase fluctuations}\label{sec:InterferenceExperiments}

The Bose--Einstein condensation is a subject of active studies
\cite{BlochLowDReview2008,ChapovskyIlichovQE2017eng,ChapovskyFedorukQE2017eng,GPforBEC2018eng,FerrariFillamentsInBEC2018eng,StringariQuanFluctAndGP2018eng,KupriyanovLightScattBEC2018eng}, which in many respects is due to the development of laser cooling and trapping~\cite{LetokhovReview2000,OnofrioCoolingReview2016,TaichenachevQE2017eng}.

\begin{figure}[htb!]
\begin{center}
\includegraphics[width=\linewidth]{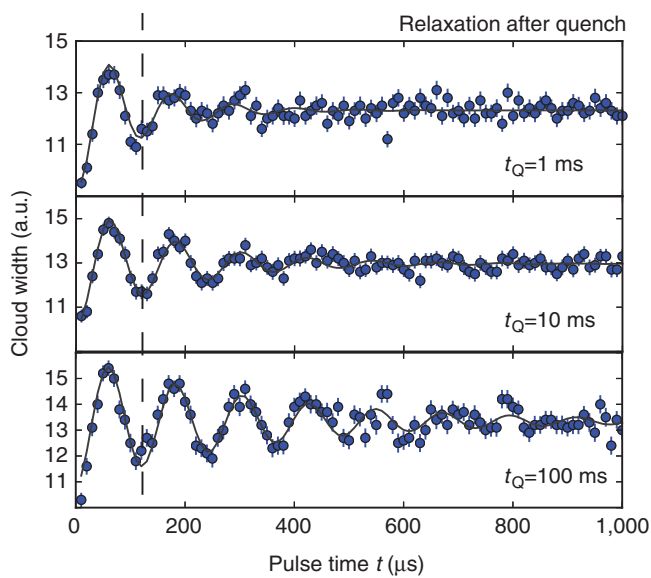}
\end{center}
\caption{FIGURE~4. The size of the expanded cloud vs the interference time $t$. A smaller value of parameter $t_{\text{Q}}$ corresponds to a preparation history that enhances fluctuations. From~\cite{FluctPhaseTalbot2017}.
}
\label{fig:OttTalbot}
\end{figure}

An influence of BEC-phase fluctuation on the Talbot effect has been observed in experiment~\cite{FluctPhaseTalbot2017}. The initial conditions are close to figure~\ref{fig:InitialCondition}a. BECs of $^{87}$Rb atoms are contained in the optical-potential minima, which are $d=547$~nm apart. At $t=0$ the confinement along $z$ is quickly extinguished, while the confinement in the $xy$ plane is left on. The BECs expand and interfere until the lattice potential is quickly reinstated at time $t$. For perfectly coherent BECs with $\varphi_j=\varphi_{j+1}$ the lattice reinstatement at $t=nT_d$ would have brought about the exact recovery of the initial state. While in the case of either $t\neq nT_d$ or $\varphi_j\neq\varphi_{j+1}$ the excited Bloch bands are partially populated, which means that the lattice reinstatement causes energy input. For detecting the energy input, the system is thermalized, then the trapping is fully turned off, and the gas expands into the free space. The introduced energy is manifested via faster expansion. The experimental dependence of the expanded cloud size on the interference time $t$ is shown in figure~\ref{fig:OttTalbot}.
The three dependencies correspond to three different histories of chain preparation. The graphs are ordered from top to bottom according to decreasing BEC phase difference. In each of the three graphs, one may see local energy minima at $t=nT_d$ corresponding to partial recovery of the initial wave function. At $t=T_d$ the experiment with smallest fluctuations (lower graph in figure~\ref{fig:OttTalbot}) corresponds to the smallest energy input in response to the lattice reinstatement.

\begin{figure}[htb!]
\begin{center}
\includegraphics[width=0.68\linewidth]{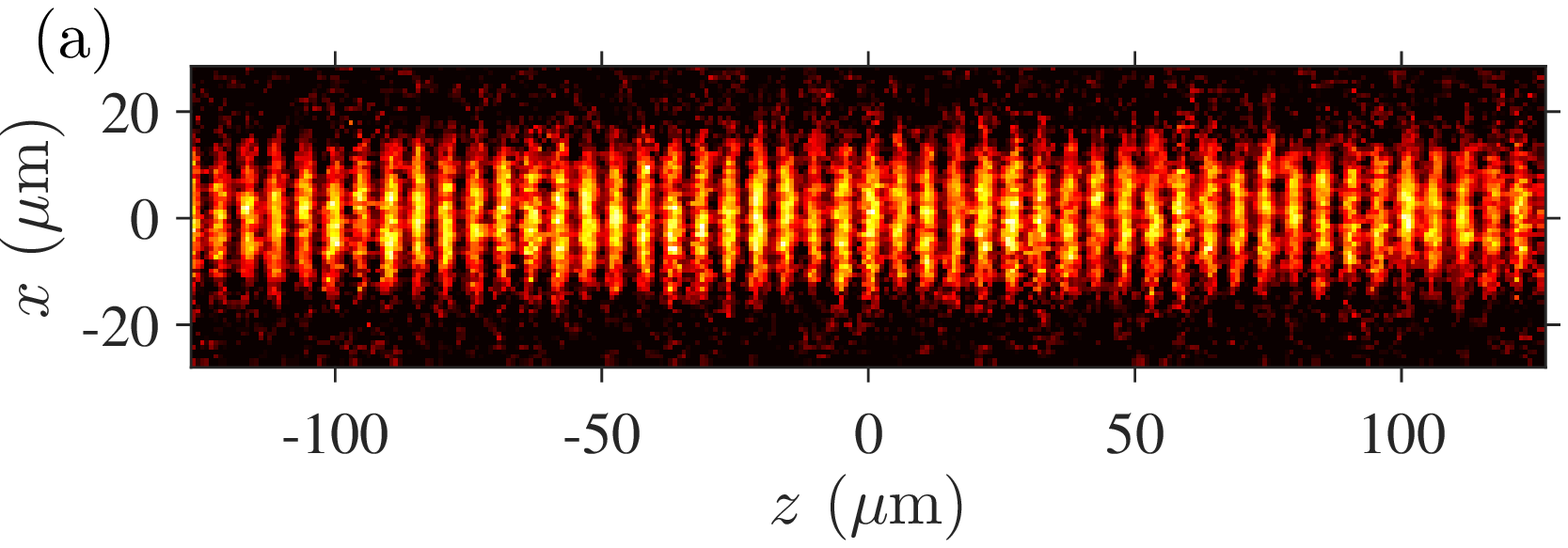}\includegraphics[width=0.3\linewidth]{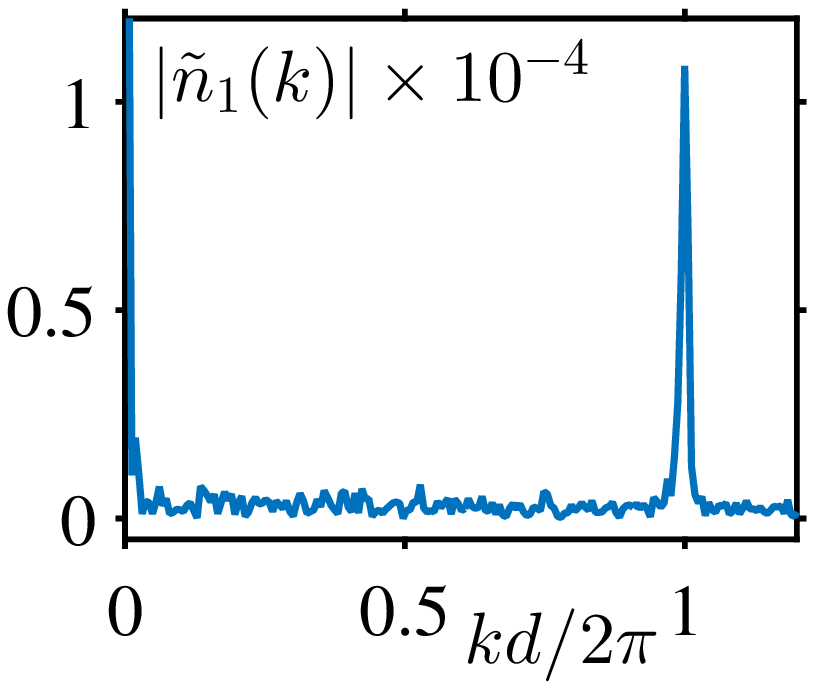}
\includegraphics[width=0.68\linewidth]{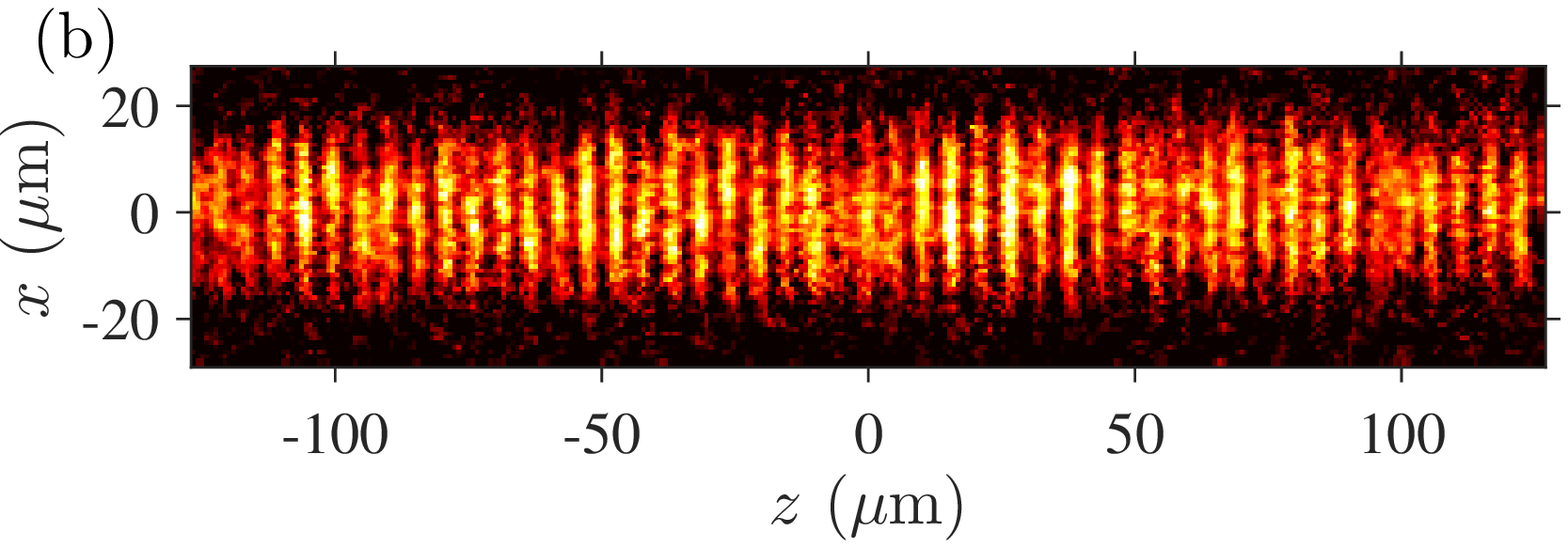}\includegraphics[width=0.3\linewidth]{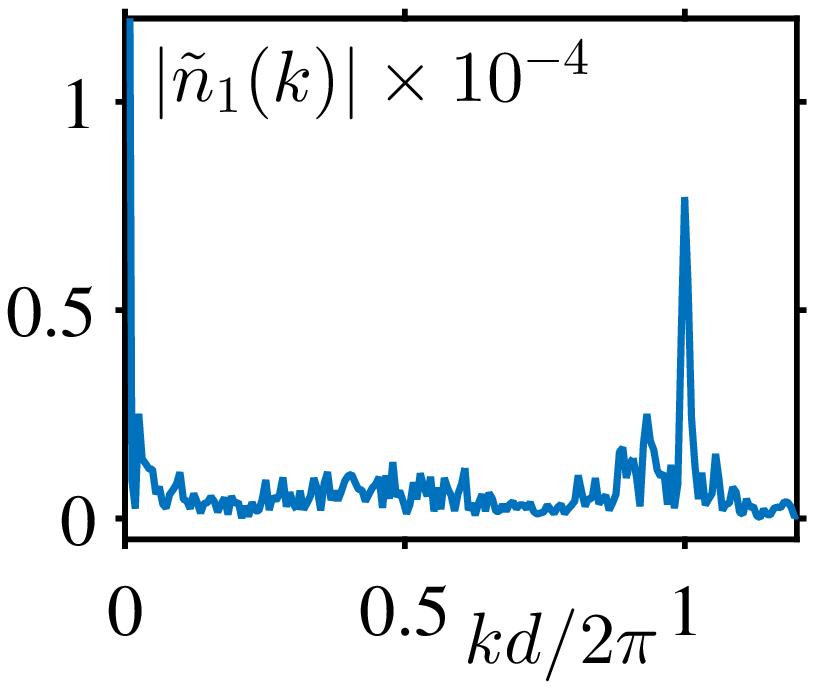}
\includegraphics[width=0.68\linewidth]{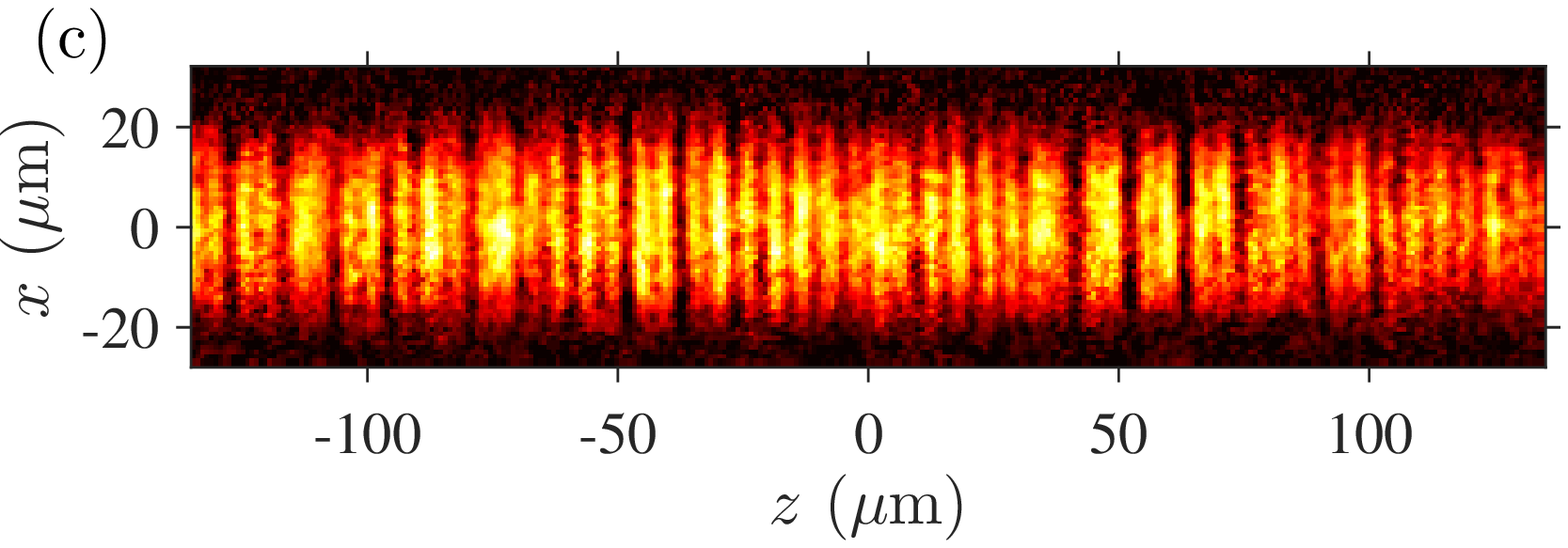}\includegraphics[width=0.3\linewidth]{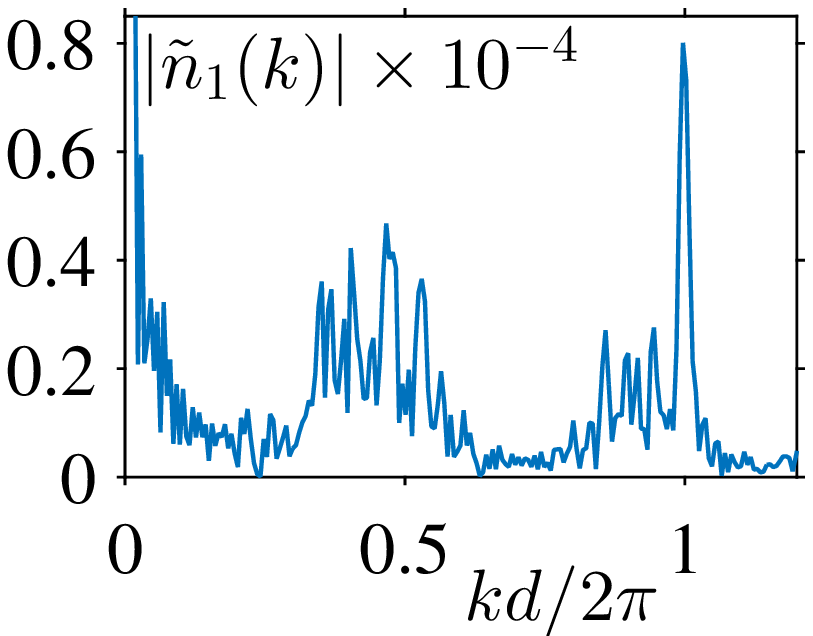}
\end{center}
\caption{FIGURE~5. Interference of a BEC chain at different instants and different correlation degree. The images are on the left, and the spatial spectra $|\tilde n_1(k)|$ are on the right.
(a) At $t=0$, as the BECs are just released from the lattice.
(b) At $t=T_d$ for strongly correlated BECs. The Talbot effect is visible.
%
(c) At $t=T_d$ for less correlated BECs. The narrow peak at $k=2\pi/d$ is a sign of the Talbot effect. The wide peaks near $k=\pi/d$ and $k=2\pi/d$ are the signs of the phase disorder.
From~\cite{RandomPhaseInteference2019}. Spectrum (c) is close to the calculated spectrum shown in figure~\ref{fig:PartialCorrSpectrum}.
}
\label{fig:TurlapovTalbot}
\end{figure}

Qualitative manifestation of phase fluctuations, which consists in the spatial spectrum alteration, is found in~\cite{RandomPhaseInteference2019}. The initial conditions correspond to figure~\ref{fig:InitialCondition}a. A chain of molecular $^6$Li$_2$ BECs is prepared in optical lattice with spacing $d=5.3$~$\mu$m between the minima. There are $N\sim1000$ bosons in each condensate. At $t=0$, the trapping field is turned off, and BECs start to expand and interfere in the free space. The interference stops at time $t$ when the system is destructively imaged. In figure~\ref{fig:TurlapovTalbot}a, the BEC chain is shown at $t=0$, when the lattice is switched off.
In spectrum $|\tilde n_1(k,t=0)|$ there is a harmonic with wave vector $k=2\pi/d$, which corresponds to the initial modulation. In figures~\ref{fig:TurlapovTalbot}b,c the system is shown at $t=T_d$. Figures~\ref{fig:TurlapovTalbot}b and \ref{fig:TurlapovTalbot}c correspond to a smaller and a bigger phase fluctuation respectively. For the small fluctuation, in figure~\ref{fig:TurlapovTalbot}b, one may see the Talbot effect. The initial distribution of the density along $z$ recovers nearly exactly. The peak $k=2\pi/d$ clearly dominates in the spectrum.

In the spectrum of figure~\ref{fig:TurlapovTalbot}c, both partial correlation and disorder are manifested. The Talbot-effect-related narrow peak at $k=2\pi/d$ is a sign of partial correlation. The broad peaks near $k=\pi/d$ and $k=2\pi/d$ are signatures of a partial phase disorder.

The spectrum in figure~\ref{fig:TurlapovTalbot}c is close to calculated $|\tilde n_1(k,T_d)|$ of figure~\ref{fig:PartialCorrSpectrum}. In experiment and calculation, for the peaks near $k=2\pi/d$ the ratio of the narrow-peak height to the wide-peak height is the same. This ratio is a function of the correlation degree $\langle\cos(\varphi_{j+1}-\varphi_j)\rangle$. Therefore, comparison of the experimental and calculational spectra yields $\langle\cos(\varphi_{j+1}-\varphi_j)\rangle$ in experiment.

The experimental spectrum in figure~\ref{fig:TurlapovTalbot}c and the calculated spectrum in figure~\ref{fig:PartialCorrSpectrum} are different in two ways. Firstly, spectrum in~\ref{fig:TurlapovTalbot}c is noisier. This is because a single-measurement outcome is shown, while the calculated $|\tilde{n}_1(k,t=T_d)|$ is averaged over 100 repetitions. Secondly, in the experiment the wide peaks are somewhat shifted to the left, which is due to the mean field of boson-boson interactions~\cite{RandomPhaseInteference2019}. The mean field is unaccounted for in the calculation.

\section{Interference-based thermometry for a chain of condensates}\label{sec:Thermometry}

The most popular method of BEC thermometry rests upon fitting condensate density profile either after release from the trap and subsequent expansion or directly in the trap. The bimodal fit yields $N_0/N$, the ratio of the condensed-particle number to the total number. The temperature may in turn be found from equation
\begin{equation}
\frac{N_0}N=1-\left(\frac T{T_{\text{c}}}\right)^D,
\end{equation}
where $D=2$ or 3 is the kinematic dimensionality and $T_{\text{c}}(N,D)$ is the Bose condensation temperature. For low $T/T_{\text{c}}$ the thermal fraction is small, and the temperature, therefore, cannot be found. For example, in \cite{OberthalerNoiseThermometry2006} the lower bound for the applicability of this method is at $T=T_{\text{c}}/2$. For two coupled condensates, a low-temperature thermometry based on their interference after the release from the trap has been demonstrated~\cite{OberthalerNoiseThermometry2006,OberthalerNoiseThermometryNJP2006}. The temperature is found from the fluctuation of the interference-fringe location in different experimental repetitions. The method requires multiple preparation of the identical pair of BECs and interference observation.

For a chain of BECs in an optical lattice the temperature may be measured in a single interference experiment. The temperature can be found from a spatial spectrum, similar to that in figure~\ref{fig:TurlapovTalbot}c. From the relation between the peaks corresponding to the Talbot effect and incoherent interference, fluctuation of adjacent condensate phases $\langle(\varphi_{j+1}-\varphi_j)^2\rangle$ may be obtained. In turn, model  \cite{BoseChainFluctPitaevskii2001} yields the connection between $\langle(\varphi_{j+1}-\varphi_j)^2\rangle$ and temperature $T$.

Model \cite{BoseChainFluctPitaevskii2001} is further applied to a chain of BECs in optical lattice with potential
\begin{equation}
V_s(\textbf{x})=sE_{\text{rec}}\left[1-e^{-2ME_{\text{rec}}(x^2+y^2)/(\hbar\lambda)^2}\cos^2\kappa z\right],
\label{eq:OptLattice}
\end{equation}
where $\textbf{x}\equiv(x,y,z)$, $\kappa$ is the optical wave vector, $E_{\text{rec}}=\hbar^2\kappa^2/2M$ is the photon recoil energy, $s$ is the dimensionless lattice depth, and $\lambda\gg1$ is the flat-cell anisotropy ratio. The period of the potential is $d=\pi/\kappa$. Near the minima, potential $V_s(\textbf{x})$ is harmonic with frequencies $\omega_z\equiv2\sqrt sE_{\text{rec}}/\hbar$ and $\omega_\perp=\omega_z/\lambda$. The description of BECs in each cell by one-particle wave functions allows to switch to Josephson Hamiltonian \cite{BoseChainFluctPitaevskii2001}
\begin{equation}
\hat H=-\frac{E_{\text{c}}}4\sum_{j=1}^{K}\frac{\partial^2}{\partial\varphi_j^2}- E_{\text{J}}\sum_{j=1}^{K-1}\cos(\varphi_{j+1}-\varphi_j).\label{eq:HJosephsonChain}
\end{equation}
Here $\hbar/E_{\text{J}}$ is the tunneling time scale, while $E_{\text{c}}=2d\mu/dN$ and $\mu$ are respectively the internal-energy scale and chemical potential of a single BEC. One may calculate the parameters of a kinematically two-dimensional BEC in lattice~(\ref{eq:OptLattice}). The chemical potential is
\begin{equation}
\mu=\hbar\omega_\perp\sqrt{2N\frac{a_{\text{b}}}{l_z}\sqrt{\frac2\pi}},
\end{equation}
where $l_z\equiv\sqrt{\hbar/(M\omega_z)}$ and $a_{\text{b}}$ is the \textit{s}-wave scattering length of the bosons. For the two-dimensional condensate, the wave function is assumed to separate into the radial and axial parts,
$\Psi_j(\textbf{x})=\psi_j(z)\Psi(\boldsymbol{\rho})$, with the radial part being the same for all BECs. This allows to write $E_{\text{J}}$ in form \cite{BoseChainExpansionPitInguscio2001}
\begin{equation}
E_{\text{J}}=\frac{\hbar^2}M \left.\left(\psi_j\frac{d\psi_{j+1}}{dz}-\psi_{j+1}\frac{d\psi_j}{dz}\right)\right|_{z=(j+1/2)d}.
\end{equation}
Under condition $\mu\ll\hbar\omega_z$, wave functions $\psi_j(z)$ may be approximated by the addends of sum (\ref{eq:PsiInitial}) with $\sigma=l_z/\sqrt2=d/(\pi\sqrt{2\sqrt s})$. Direct calculation yields
\begin{equation}
E_{\text{J}}=\frac{\hbar^2}M\frac{Nde^{-d^2/8\sigma^2}}{\sigma^32\sqrt{2\pi}}= \frac{\hbar^2}{Md^2}N\pi^2\sqrt\pi\sqrt{s\sqrt s}e^{-\pi^2\sqrt s/4}.
\end{equation}
It is worth of noting that the use of Gaussian functions of width $\sigma$ narrows down the BEC wave function and, therefore, gives lower bound of $E_{\text{J}}$, which approaches the true value with increasing depth $s$.

Consider the case of small quantum phase fluctuations. The quantum part of the fluctuations is computed from Hamiltonian (\ref{eq:HJosephsonChain}) at $T=0$. This gives the condition of the quantum-fluctuation smallness
\begin{equation}
\langle(\varphi_{j+1}-\varphi_j)^2\rangle=\frac12\sqrt{\frac{E_{\text{c}}}{E_{\text{J}}}}\ll1.
\end{equation}
Also assume that the thermal fluctuations dominate. In this case \cite{BoseChainFluctPitaevskii2001}
\begin{equation}
\langle\cos(\varphi_j-\varphi_l)\rangle=\alpha^{|j-l|} \label{eq:ThermalFluct}
\end{equation}
with $\alpha$ being function of temperature \cite{BoseChainFluctPitaevskii2001,OberthalerNoiseThermometryNJP2006}
\begin{equation}
\alpha(T)= 
\frac{I_1(E_{\text{J}}/T)}{I_0(E_{\text{J}}/T)},
\end{equation}
where $I_\nu$ is the modified Bessel function. Expression (\ref{eq:ThermalFluct}) simplifies for $T\ll E_{\text{J}}$:
\begin{equation}
\langle(\varphi_j-\varphi_l)^2\rangle=|j-l|\frac T{E_{J}}.\label{eq:SmallThermalFluct}
\end{equation}
The connection between the temperature and the BEC-phase fluctuations is established by expression (\ref{eq:ThermalFluct}) and its particular case (\ref{eq:SmallThermalFluct}). Therefore, formulas (\ref{eq:ThermalFluct}) and (\ref{eq:SmallThermalFluct}) together with the spectrum analysis may serve as a basis for thermometry.

\section{Conclusion}\label{sec:Concl}

Partial phase disagreement between the sources alters the Talbot effect. Quantitatively this is manifested by a decrease of the interference contrast. Qualitatively the change is represented by new peaks in the spatial spectrum of the interference fringes. This change of the spectrum may be used for thermometry.

V.~B.~Makhalov is thankful for financial support to the Russian foundation for basic research, grants 15-02-08464 and 18-42-520024. A.~V.~Turlapov is thankful to Russian science foundation, grant 18-12-00002.

\end{document}